\documentclass[10pt,twoside,BCOR7mm,DIV15,headinclude,footexclude,cleardoubleempty,idxtotoc]{scrartcl}

\usepackage[english]{babel}
\usepackage{graphicx}
\usepackage{hyperref}
\usepackage{scrpage2}
\usepackage{hyperref}
\usepackage{ifthen}

\makeatletter
\renewcommand{\@biblabel}[1]{}
\renewcommand{\@cite}[2]{%
{#1\ifthenelse{\boolean{@tempswa}}{,#2}{}}}
\makeatother

\hypersetup{breaklinks=true,colorlinks=true,linkcolor=black,urlcolor=blue,citecolor=blue}

\ofoot{\thepage}
\ifoot{}

\setheadsepline{1pt}

\setkomafont{pagehead}{\normalfont\sffamily}
\setkomafont{pagenumber}{\normalfont\rmfamily}

\usepackage{booktabs}
\usepackage{amsmath}
\usepackage{amssymb}
\usepackage{multicol}
\usepackage{float}

\makeatletter
\newcommand{\listofcontributions}{\@starttoc{con}}

\newcommand{\l@contribution} {\@dottedtocline{1}{1.5em}{2.3em}}
\makeatother

\newenvironment{contribution}{
\setcounter{section}{0}
\setcounter{figure}{0}
\setcounter{table}{0}
\begin{flushleft}
{\em Clumping in Hot Star Winds \\
W.-R.\ Hamann, A.\ Feldmeier \& L.\ Oskinova, eds.\\
Potsdam: Univ.-Verl., 2007 \\
URN: http://nbn-resolving.de/urn:nbn:de:kobv:517-opus-13981
} 
\end{flushleft}
}{
\newpage
\lehead{}
\rohead{}
}

\begin{document}

\setlength{\baselineskip}{2.5ex}

\begin{contribution}
%%% 3 pages !!!!!
%%%%%%%%%%%%%%%%%%%%%%%%%%%%%%%%%%%%%%%%%%%%%%%%%%%%%%%%%%%%%%%%%%%%%%%
% EXAMPLE AND TEMPLATE FILE FOR PROCEEDINGS OF THE CLUMPING WORKSHOP.
% PLEASE REPLACE THE TEMPLATE TEXT BY YOUR OWN ARTICLE.
% NOTE THAT YOU MUST NOT PROCESS THIS FILE, BUT THE MASTER FILE:
% latex masterfile; dvips masterfile

\newcommand\sun{\odot}
\newcommand\arcdegr{\mbox{$^\circ$}} 
\newcommand\arcmin{\mbox{$^\prime$}}

\newcommand\aj{{AJ}}% Astronomical Journal 
\newcommand\araa{{ARA\&A}}% Annual Review of Astron and Astrophys 
\newcommand\apj{{ApJ}}% Astrophysical Journal 
\newcommand\apjl{{ApJ}}% Astrophysical Journal, Letters 
\newcommand\apjs{{ApJS}}% Astrophysical Journal, Supplement 
\newcommand\apss{{Ap\&SS}}% Astrophysics and Space Science 
\newcommand\aap{{A\&A}}% Astronomy and Astrophysics 
\newcommand\aapr{{A\&A~Rev.}}% Astronomy and Astrophysics Reviews 
\newcommand\aaps{{A\&AS}}% Astronomy and Astrophysics, Supplement 
\newcommand\baas{{BAAS}}% Bulletin of the AAS 
\newcommand\mnras{{MNRAS}}% Monthly Notices of the RAS 
\newcommand\pasp{{PASP}}% Publications of the ASP 
\newcommand\pasj{{PASJ}}% Publications of the ASJ 
\newcommand\ssr{{Space~Sci.~Rev.}}%          % Space Science Reviews  
\newcommand\nat{{Nature}}% Nature 
\newcommand\iaucirc{{IAU~Circ.}}% IAU Cirulars 
\newcommand\aplett{{Astrophys.~Lett.}}% Astrophysics Letters 
\newcommand\apspr{{Astrophys.~Space~Phys.~Res.}}% Astrophysics Space Physics Research 

% RUNNING AUTHOR: PUT AUTHOR NAMED HERE
\lehead{O.\ Reimer, F.\ Aharonian, J.\ Hinton, W.\ Hofmann, S.\ Hoppe, M.\ Raue \& A.\ Reimer}

% RUNNING TITLE; SHORTEN THE TITLE IF NECESSARY
% IN CASE OF A ONE-PAGE CONTRIBUTION (POSTER),
% SQUEEZE AUTHORS AND TITLE IN THIS LINE (Author: Title ...)
\rohead{VHE gamma-rays from Westerlund~2}

\begin{center}
% FULL TITLE HEADING
{\LARGE \bf VHE gamma-rays from Westerlund~2 and implications for the inferred energetics}\\
\medskip

% AUTHORS LIST
{\it\bf O.~Reimer$^1$, F.~Aharonian$^{2,3}$, J.~Hinton$^4$, W.~Hofmann$^2$,\\ S.~Hoppe$^2$, M.~Raue$^5$, and A.~Reimer$^1$}\\

% AFFILIATIONS
{\it $^1$ W.W. Hansen Experimental Physics Laboratory \& \\ Kavli Institute for Particle Astrophysics and Cosmology, Stanford University, USA}\\
{\it $^2$ Max-Planck-Institut f\"ur Kernphysik, Heidelberg, Germany}\\
{\it $^3$ Dublin Institute for Advanced Studies, Ireland}\\
{\it $^4$ School of Physics \& Astronomy, University of Leeds, UK }\\
{\it $^5$ Institut f\"ur Experimentalphysik, Universit\"at Hamburg, Germany}\\

% ABSTRACT
\begin{abstract}

The H.E.S.S. collaboration recently reported the discovery of VHE $\gamma$-ray emission coincident with 
the young stellar cluster Westerlund~2. This system is known to host a population of hot, massive stars, 
and, most particularly, the WR binary WR~20a.  Particle acceleration to TeV energies in Westerlund~2 can be 
accomplished in several alternative scenarios, therefore we only discuss energetic constraints based on the 
total available kinetic energy in the system, the actual mass loss rates of respective cluster members, 
and implied gamma-ray production from processes such as inverse Compton scattering or neutral pion decay. 
From the inferred gamma-ray luminosity of the order of $10^{35}$erg/s, implications for the efficiency of 
converting available kinetic energy into non-thermal radiation associated with stellar winds in the Westerlund~2 
cluster are discussed under consideration of either the presence or absence of wind clumping.
\end{abstract}
\end{center}

\begin{multicols}{2}

\section{The stellar cluster Westerlund~2 in the HII region RCW~49}
The prominent giant HII region RCW~49 is characterized by still ongoing massive star formation (\cite{ref1}).
The regions surrounding the central stellar cluster Westerlund~2 appear evacuated by stellar winds and radiation, 
and dust is distributed in fine filaments, knots, pillars, bubbles, and bow shocks throughout the rest of the HII complex (\cite{ref2, ref3}). 
Radio continuum observations revealed two wind-blown shells in the core of RCW~49 (\cite{ref4}), surrounding the central region of Westerlund~2, 
and the prominent Wolf-Rayet star WR~20b. There is an ongoing controversy over the distance to Westerlund~2, 
and consequently about the association of WR~20a with Westerlund~2, as will be discussed later. The stellar cluster 
contains an extraordinary ensemble of hot and massive stars, at least a dozen early-type O-stars, and two remarkable WR stars. 
One of them, WR~20a was only recently established to be a binary (\cite{ref5, ref6}) by presenting solutions for a circular orbit 
with a period of 3.675, and 3.686 days, respectively. Based on the orbital period, the minimum masses were found to be 
$(83 \pm 5)$\,M$_{\odot}$ and $(82 \pm 5)$\,M$_{\odot}$ for the binary components (\cite{ref7}). At that time, WR~20a was classified 
as the most massive of all confidently measured binary systems in our Galaxy. Synchrotron emission has not yet been detected 
from the WR~20a system, presumably because of free-free-absorption in the optically thick stellar winds along 
the line of sight. Although WR~20a has been detected in X-rays (\cite{ref8}), the non-thermal and thermal components of the X-ray 
emission remain currently indistinguishable. Detectable VHE gamma-radiation from the WR~20a binary system was only predicted 
in a pair cascade model (\cite{ref9}), although detailed modeling of the WR~20a system in other scenarios 
(e.g. as of \cite{ref10} when produced either by optically-thin inverse Compton scattering of relativistic electrons with the 
dense photospheric stellar radiation fields in the wind-wind collision zone or in neutral pion decays, with the mesons produced 
by inelastic interactions of relativistic nucleons with the wind material) is still pending. In VHE $\gamma$-rays, photon-photon 
absorption would modulate (and diminish) the observable flux from a close binary system such as WR~20a. 

%-----------Double-column figure -----------------------------------
\begin{figure*}[!t]
\begin{center}
\includegraphics[width=0.90\textwidth]{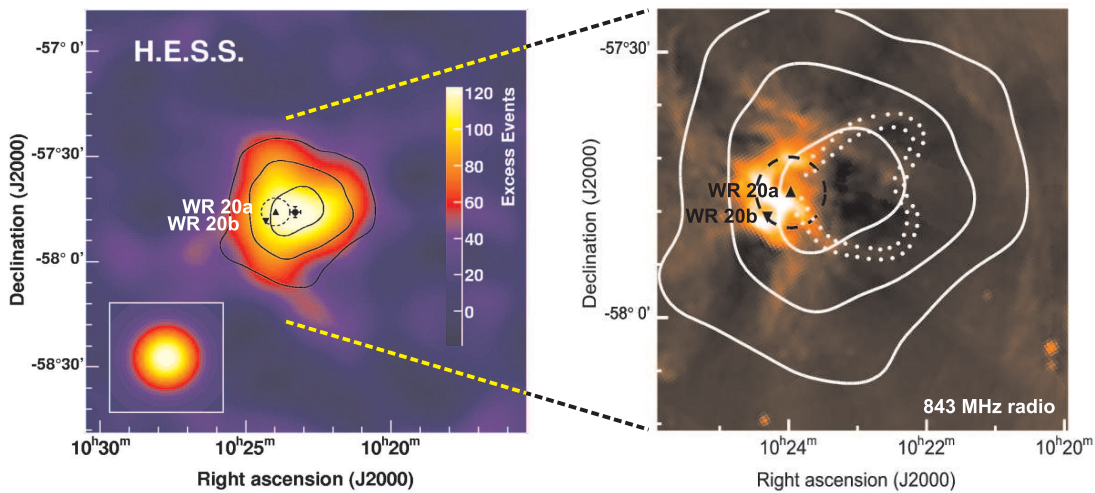}
\caption{Left: H.E.S.S. gamma-ray sky map of the Westerlund 2 region, smoothed to reduce the effect of statistical fluctuations. 
The inlay in the lower left corner shows how a point-like source would have been seen by H.E.S.S. The WR stars WR~20a and WR~20b 
are marked as triangles, and the stellar cluster Westerlund 2 is represented by a dashed circle.  Right: Significance contours of 
the gamma-ray source HESS~J1023--575 (corresponding 5, 7 and 9$\sigma$), overlaid on a radio image from the Molonglo Observatory 
Synthesis Telescope. The wind-blown bubble around WR~20a, and the blister to the west of are seen as depressions in the radio continuum. 
The \emph{blister} is indicated by white dots, and appears to be compatible in direction and location with HESS~J1023--575.
\label{OReimer:Fig1}}
\end{center}
\end{figure*}
%----------------------------------------------------------------

\section{H.E.S.S. observations towards Westerlund~2}

The H.E.S.S. (High Energy Stereoscopic System) collaboration observed the Westerlund~2 region between March and July 2006, 
and obtained 14 h (12.9 h live time) of data, incorporating targeted observations of WR~20a and data from the ongoing H.E.S.S. 
Galactic plane survey. The data were obtained under zenith angles in the range between 36$^\circ$ and 53$^\circ$, 
resulting in an energy threshold of 380 GeV for the analysis. A point source analysis on the nominal position of WR~20a resulted 
in a clear signal with a significance of 6.8$\sigma$, and further investigations revealed an \emph{extended} excess with a peak 
significance exceeding 9$\sigma$. The center of the excess was derived by fitting the two-dimensional point spread function of the instrument folded with a Gaussian to the uncorrelated excess map:  
$\alpha_{2000}$ = $10^{\rm h}23^{\rm m}18^{\rm s} \pm 12^{\rm s}$, $\delta_{2000}$ = -57$^\circ$45'50'' $\pm$ 1'30''. 
The systematic error in the source location is 20'' in both coordinates. The source is clearly extended beyond the appearance
of a point-like source for the H.E.S.S. instrument (Fig.~1), and a fit of a Gaussian folded with the PSF gives an rms extension of 
$0.18^\circ \pm 0.02^\circ$. The differential energy spectrum can be described by a power law  
dN/dE$= \Phi_0 \cdot (\mbox{E}/1\,\mbox{TeV})^{- \Gamma}$ with a photon index of 
$\Gamma=2.53 \pm 0.16_{\mathrm{stat}} \pm 0.1_{\mathrm{syst}}$ and a normalization at 1\,TeV of 
$\Phi_0 = (4.50 \pm 0.56_{\mathrm{stat}} \pm 0.90_{\mathrm{syst}}) \times 10^{-12}$\,TeV$^{-1}$\,cm$^{-2}$\,s$^{-1}$. 
The integral flux for the whole excess above the energy threshold of 380 GeV is (1.3 $\pm$ 0.3) $\times 10^{-11}$\,cm$^{-2}$\,s$^{-1}$. 
No significant flux variability or the characteristic orbital periodicity of WR~20a could be detected in the data set. Full details
regarding the discovery of HESS~J1023--575 at VHE $\gamma$-rays are given in \cite{ref11}.

\section{Size constraints and energetics}

The detection of extended VHE $\gamma$-ray emission towards Westerlund~2 is indicative of the presence of extreme high-energy 
particle acceleration in this young ($\sim$2-3 Myrs;~\cite{ref12}) star forming region. Following the HEGRA detected source 
TeV~J2032+4130 and its suggested connection to the Cygnus OB2 cluster (\cite{ref13}), HESS~J1023--575 and 
Westerlund~2 is the second but even more prominent association between VHE $\gamma$-ray emission and an extraordinary 
assembly of young, hot and massive stars in our Galaxy. Given that the size of the $\gamma$-ray emission does not resemble 
the nominal size of the stellar cluster as known from radio, infrared to optical, and X-ray energies very well, but stretches 
further out in the direction of the \emph{blister} (\cite{ref4}), we discuss the implied energetics based on the most simple possible, 
and accordingly least model-dependent considerations.  A central problem for any stringent energetic assessment lies in the still unsettled 
dispute on the distance to Westerlund~2, when even recent determinations differ apparently by more than a 
factor of 3 (see Fig.~2): The distance to Westerlund~2 is uncertain in the range of values between $\sim$2.2 kpc~(\cite{ref15}) and 
7.9 kpc~(\cite{ref16}), and intermediate values of 4.2 kpc were derived from 21 cm absorption line profile measurements~(\cite{ref17}), 
5.75 kpc from the distance estimate towards the prominent WR star WR~20a~(\cite{ref18}), and 6.4 kpc from photometric 
measurements~(\cite{ref19}). Recently, Rauw et al. (2007) presented a compelling re-determination of the distance to Westerlund~2 by 
spectro-photometric measurements of 12 cluster member O-type stars of (8.3$\pm$1.6)~kpc, a value in very good agreement with the 
(8.0$\pm$1.0)~kpc as measured by \cite{ref7} as determined from the light curve of the eclipsing binary WR~20a. We adopt the value 
of the weighted mean of (8.0$\pm$1.4)~kpc (\cite{ref20}) throughout this manuscript, thereby associating WR~20a as a cluster member 
of Westerlund~2. Note, however, that \cite{ref21} and \cite{ref22} put forward significantly lower values.   

%-----------Double-column figure -----------------------------------
\begin{figure}[H]
\begin{center}
\includegraphics[width=0.48\textwidth]{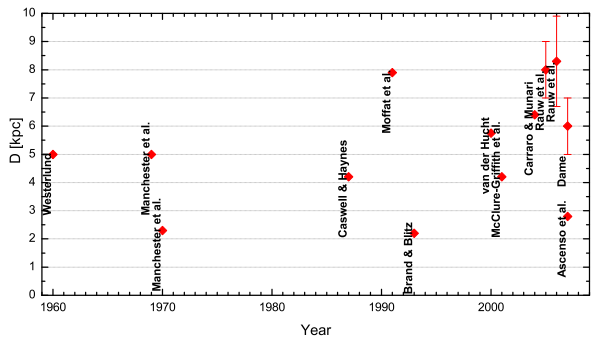}
\caption{Distance measures for Westerlund~2 and/or WR~20a. The lasting distance ambiguity from different 
methodological approaches translates into the energetic constraints imposed by the detection of VHE 
$\gamma$-radiation. 
\label{OReimer:Fig2}}
\end{center}
\end{figure}
%-----------------------------------------------------------
With a projected angular size of submilliarcsecond scale, the WR~20a binary system, including its colliding wind zone, 
would appear as a point source for observations with the H.E.S.S. telescope array. At a distance of 8 kpc, the measured 
$\gamma$-ray source extension is equivalent to a diameter of 28 pc. Unless there are extreme differences in the spatial 
extent of the particle distributions producing radio, X-ray, and VHE $\gamma$-ray emission, scenarios based solely on 
particle acceleration in the colliding wind zone of WR~20a are unlikely to account for the observed source extent of 
0.18$^\circ$ in the VHE $\gamma$-rays. Therefore the bulk of the $\gamma$-rays cannot energetized to TeV energies close 
to WR~20a. The apparent size of the VHE photons is however consistent with 
theoretical predictions of bubbles blown from massive stars into the ISM~(\cite{ref23}).  

We estimated the $\gamma$-ray luminosity above 380 GeV to $\sim 1.5\times 10^{35}$erg/s (at 8~kpc), corresponding to 0.2\% 
(smooth wind: $\dot M$ = 2.5$\times 10^{-5} M_\sun$/yr) or 0.7\% (clumped wind: $\dot M$ = 8.5$\times 10^{-6} M_\sun$/yr) 
of the total kinetic energy available from the colliding winds of WR~20a, and 0.2\% (smooth wind: $\dot M$ = 5.3$\times 10^{-5} M_\sun$/yr) 
or 0.7\% (clumped wind: $\dot M$ = 1.7$\times 10^{-5} M_\sun$/yr) of the kinetic energy of WR~20b, respectively. With up to 1.4\% of the 
available $E_{kin}$ in the WR-winds alone, a canonical value of 10\% acceleration efficiency, the implied $\gamma$-ray production 
efficiency ($L_{\gamma,\mathrm{VHE}}/L_\mathrm{particle}$) is as high as 14\% in case of clumped winds, or 4\% for the less realistic case of
smooth winds. These estimates, however, do not consider additional mass loss from other hot and massive stars present in Westerlund~2,  
which needs to be included when the H.E.S.S. result is interpreted in terms of collective stellar wind outflows. The energetic 
constraint is further relaxed when the distance to Westerlund~2 is indeed lower than the 8 kpc assumed, due to the accordingly higher integral 
$\gamma$-ray luminosity, and lower efficiency $L_{\gamma,\mathrm{VHE}}/L_\mathrm{particle}$.\\ 
In summary, the inferred $\gamma$-ray luminosity from the detection of VHE $\gamma$-ray emission from Westerlund~2 implies a rather 
high conversion efficiency when considering the mass loss rates for clumped winds from the two WR stars. This constraint 
can be relaxed if Westerlund~2 is indeed more closer than 8 kpc, as well as by considering mass loss transfer from the O- and B-stars in the stellar cluster.\\  

% REFERENCES IN ALPHABETICAL ORDER

\end{multicols}

\end{contribution}

%%-------------------------------------------------------

\end{document}